# GLOBAL AI GOVERNANCE: WHERE THE CHALLENGE IS THE SOLUTION
# - AN INTERDISCIPLINARY, MULTILATERAL, AND VERTICALLY COORDINATED APPROACH *

**Huixin Zhong**
Xi'an Jiaotong Liverpool University
University of Bath
Huixin.Zhong@XJTLU.edu.cn

**Thao Do**
University of Bath
tnd32@bath.ac.uk

**Yangliu Jie**
ESCP Business School
yangliujie.eu@gmail.com

**Rostam J. Neuwirth**
Faculty of Law
Department of Global Legal Studies
University of Macau
rjn@um.edu.mo

**Hong Shen**
Carnegie Mellon University
hongs@andrew.cmu.edu

## ABSTRACT

Current global AI governance frameworks struggle with fragmented disciplinary collaboration, ineffective multilateral coordination, and disconnects between policy design and grassroots implementation. This study, guided by Integration and Implementation Science (IIS) initiated a structured interdisciplinary dialogue at the UN Science Summit, convening legal, NGO, and HCI experts to tackle those challenges. Drawing on the common ground of the experts: dynamism, experimentation, inclusivity, and paradoxical governance, this study, through thematic analysis and interdisciplinary comparison analysis, identifies four core principles of global AI governance. Furthermore, we translate these abstract principles into concrete action plans leveraging the distinct yet complementary perspectives of each discipline. These principles and action plans are then integrated into a five-phase, time-sequential framework including foundation building, experimental verification, collaborative optimization, global adaptation, and continuous evolution phases. This multilevel framework offers a novel and concrete pathway toward establishing interdisciplinary, multilateral, and vertically coordinated AI governance, transforming global AI governance challenges into opportunities for political actions.

## 1 Introduction

The rapid progress of artificial intelligence (AI) has underscored the urgent need for effective regulatory oversight. As a result, major powers like the EU [1], the US [2], China [3], and the UK [4] have developed distinct approaches to governing AI. This has created a competitive landscape, often described as a "regulatory race," or even "regulatory battle" where these powers vie to shape global AI norms and standards [5], [6], [7].

However, this race-based narrative has been criticized as ineffective in addressing the global challenges of AI [8], [9]. AI's global impact, as emphasized by UNESCO [10], necessitates a unified regulatory framework. The transnational nature of AI, where algorithms, data, and systems operate across borders, makes it difficult for individual nations to control cross-border flows and technology transfer, leading to regulatory fragmentation [11], [12]. Incompatible national rules create compliance challenges, enabling regulatory arbitrage that undermines global governance efforts.

---

*\**Paper under review*

In response, international organizations like the United Nations are working to establish a unified AI governance framework [13], [14]. However, challenges persist, including the lack of a recognized authority for comprehensive and interdisciplinary AI assessments [13], complex multilateral engagement [15], and insufficient coordination between top-down and bottom-up approaches [16].

This study tackles the above challenges of global AI governance through the Integration and Implementation Sciences (IIS). IIS offers a comprehensive approach to addressing complex global issues by integrating research, policy, and practice across multiple domains [17]. Its emphasis on multi-stakeholder collaboration, systems thinking, and knowledge management [17] makes it particularly suitable for navigating the complexities of global AI governance. By convening a diverse panel of experts, we operationalized IIS principles to uncover governance patterns and reconcile cross-domain perspectives. This methodological approach led to the development of a multilevel global AI governance framework.

This multilevel framework integrates four core principles: dynamism, experimentation, inclusivity, and paradoxical governance—within a five-phase, time-sequential process: experimental verification, collaborative optimization, global adaptation, and continuous evolution. Our study makes three key contributions. Theoretically, it introduces a novel framework to address critical gaps in global AI governance. Methodologically, it pioneers the application of Integration and Implementation Science (IIS) to translate AI governance challenges into actionable solutions. Practically, it constructs concrete implementation pathways, based on interdisciplinary complementarity, across macro, meso, and micro levels. This enables governance solutions to smoothly transition from abstract high-level principles to practical and effective management measures.

The subsequent sections begin by diagnosing the current gaps in global AI governance, followed by an introduction to the research methods grounded in Integration and Implementation Sciences (IIS). We then present expert perspectives through thematic analysis, which is further enriched by an interdisciplinary comparative analysis to extract core principles. Finally, we propose a synthesized framework based on these principles, structured as a chronological five-phase implementation roadmap.

## 2 Literature Survey

### 2.1 Universal Challenges in Global AI Governance

#### 2.1.1 Lack of Effective Interdisciplinary Collaboration

Whether establishing international standards or formulating national policies, a comprehensive assessment of AI systems' impacts and future trajectories is essential. However, effectively integrating expertise from diverse disciplines remains a significant hurdle [18]. Scientific research is often highly specialized, with distinct methodologies and terminologies across different fields, making interdisciplinary collaboration exceptionally challenging. For instance, establishing effective communication and collaboration mechanisms among fields like ethics, computer science, law, and social sciences requires overcoming substantial differences in knowledge and methodologies. Equally, AI posing a highly interdisciplinary and complex problem is illustrated by challenges posed by the regulation of a number of prohibited AI systems, such as notably those capable of manipulating the human mind [19], [20]. Many national AI strategies or ethical guidelines, while emphasizing the importance of interdisciplinary collaboration, often lack concrete implementation plans and mechanisms (e.g., Government of Canada, 2017; Council for Science, Technology and Innovation, 2019). This "lip service" reflects the universal challenge of interdisciplinary collaboration on a global scale.

#### 2.1.2 The Dilemmas of Multilateralism

While many nations recognise the necessity of international cooperation in AI governance, multilateralism faces numerous practical challenges. Diverging national interests, geopolitical competition, and concerns about data sovereignty can all hinder the achievement of international consensus. For example, significant differences in national positions exist on key issues such as cross-border data flows, algorithmic transparency, and the weaponization of AI, leading to slow progress in international cooperation ([15]). A major concern in this respires is a highly fragmented international institutional system that is in urgent need of reform [8]. Even at the regional level, such as with the EU's AI Act, negotiations among member states on specific provisions and enforcement mechanisms face complexities (Bradshaw et al., 2021). This dilemma of multilateralism reflects the universal challenge of achieving international cooperation while safeguarding national interests in a globalized world.

### 2.1.3 Insufficient Top-Down and Bottom-Up Coordination

Existing AI governance models often lean towards top-down approaches, with policy formulation and implementation primarily led by governments, lacking sufficient participation from grassroots stakeholders ([16]). This can lead to policies disconnected from practical needs and fail to adequately consider the demands of ordinary users, small businesses, and civil society organizations. For example, some countries' implementation of facial recognition technology regulations, lacking adequate communication and consultation with the public, has raised concerns about privacy violations and social surveillance (Zuboff, 2019). Furthermore, large technology companies, wielding significant influence and resources in the AI field, often dominate policy discussions, while the interests of smaller businesses and startups are easily overlooked ([21]). This issue of insufficient top-down and bottom-up coordination is a global phenomenon, hindering fair, inclusive, and effective AI governance.

These challenges are not isolated but rather interconnected and mutually reinforcing. A lack of interdisciplinary collaboration weakens the foundation of multilateralism, while insufficient top-down and bottom-up coordination exacerbates the dilemmas of multilateralism. Therefore, constructing an effective global AI governance framework requires fundamentally addressing these universal challenges, promoting deep interdisciplinary integration, strengthening multilateral dialogue and cooperation, and establishing more inclusive and participatory governance models.

## 3 Research Method

To address the complex challenges of global AI governance, we adopted the transdisciplinary paradigm of Integration and Implementation Sciences (IIS) [17], a comprehensive approach that systematically links research, policy, and practice through three core pillars: multi-stakeholder collaboration, systems thinking, and knowledge management. This IIS orientation shaped both the organizational design of our initiative and our analytical strategy, enabling an integrative approach to complex governance issues.

### 3.1 Event Convention and Strategic Sampling

In line with IIS's participatory methods and knowledge-sharing principles [17], which emphasize stakeholder involvement and the integration of diverse perspectives to enhance the acceptance, effectiveness, and sustainability of governance frameworks, we convened a diverse expert panel at the United Nations Science Summit. This panel served as a participatory platform to facilitate inclusive dialogue and collaborative knowledge production, central to IIS's approach.

To ensure meaningful representation and capture the multifaceted nature of global AI governance, we adopted a strategic sampling approach grounded in the principle of "minimum necessary diversity" [22]. Experts were purposefully selected from three key domains—law, Global South NGOs, and human-computer interaction (HCI)—reflecting critical dimensions of AI governance. This sampling design was based on the 'authority-justice-technology' tension triangle of AI governance [23], aims to capture: (a) the macro-level tension of institutional authority (the legal expert); (b) the meso-level dilemma of distributive justice (the global south NGO expert); and (c) the micro-level challenge of technological democratization (the HCI expert). Together, these perspectives providing a robust foundation for interdisciplinary analysis and framework development.

### 3.2 Data Collection and Analysis

Our data analysis combined two interrelated strategies. First, we conducted systematic thematic analysis of expert inputs, reflecting IIS's knowledge management pillar by identifying governance priorities through iterative coding of multi-stakeholder perspectives. Second, we applied IIS-informed interdisciplinary comparative analysis to identify both convergences and productive tensions across disciplines, leveraging systems thinking to generate interdisciplinary insights.

Data were collected at the UN Science Summit in two stages: domain-specific presentations (20 minutes each) from experts in law, NGOs, and HCI, followed by panel discussions (30 minutes) exploring practical governance paradoxes within and across these fields.

Thematic analysis [24], and interdisciplinary comparative analysis [25], involved three phases. First, two researchers independently coded presentation transcripts, identifying key sub-themes (15 sub-themes of the legal expert, 14 of the NGO practitioner, 11 of the HCI expert. The initial transcript and detailed thematic analysis are available in Open Science Framework (OSF) via the anonymous link: https://osf.io/qm7zv/?view_only=66700c671a144dd29e565e96df0f0b1a). These sub-themes were clustered into three meta-propositions within each domain. Following Repko's interdisciplinary transformation strategy [25], an interdisciplinary comparative analysis then revealed four core principles: dynamism, experimentation, inclusivity, and paradox management based on the common ground of three experts. Finally, these abstract principles were translated into action plans, informed by the distinct yet complementary perspectives of three domain experts, and embedded within a five-phase, multilevel AI governance framework detailed in section six.

## 4 Thematic Analysis: An Interdisciplinary Perspective of Global AI Governance

We first analyzed the keynote speeches and panel discussions of each speaker using thematic analysis. Thematic analysis is well-suited for this study because it helps us break down complex ideas into manageable themes, making it easier to understand how each discipline—law, NGO, and HCI—contributes to the overall framework. For each expert, three main themes were concluded.

### 4.1 Perspective of Legal Expert

#### 4.1.1 Interdisciplinary Convergence for Cognitive Frameworks

Legal experts point out that AI both reflects and challenges human cognitive intelligence, and the over-specialization of disciplines leads to knowledge fragmentation [26]. It is necessary to establish interdisciplinary dialogue mechanisms to promote the deep integration of linguistics, ethics, and computer science, and to achieve substantive integration of cognitive frameworks beyond the translation of terms. This integration is not a simple superposition of knowledge, but the construction of a common language to understand the principles of human cognition and the relationship between AI, laying the foundation for addressing ethical, social, and technical challenges.

#### 4.1.2 Paradox Navigation in Global Governance

This theme reveals a deep paradox in global AI governance: international institutions, while providing legitimacy, are constrained by structural inefficiencies and conflicts of interest, creating a cyclical dilemma of "institutions first or practice-driven" (a "chicken and egg paradox") [27]. Rosenau and Schinagl point out that this governance complexity and the borderless nature of cyberspace require dynamic response mechanisms [28], [29]. Key strategies include: (1) leveraging AI and communication technology innovations to develop decentralized governance tools, breaking through traditional institutional constraints; (2) constructing regulatory frameworks with dual temporal and spatial dimensions—establishing continuous adaptation mechanisms in time to match the speed of technological iteration, and designing cross-border collaboration solutions in space to coordinate jurisdictional conflicts. These measures aim to transform governance paradoxes into evolutionary driving forces, providing flexible solutions for transnational issues.

#### 4.1.3 Ethical, Societal, and Cognitive Transformations of AI: Embracing Complexity

AI is driving three paradigm shifts. Ethically, AI presents dilemmas such as algorithmic bias, autonomous weapons, privacy erosion, and labor market impacts, requiring a balance between seemingly conflicting values like data use vs. privacy and efficiency vs. human autonomy. Addressing these challenges demands a nuanced approach beyond simplistic solutions. Socially, AI reshapes communication, information access, and social relationships, with the potential to both enhance connectivity and widen inequalities. Solutions must promote inclusivity, fairness, and social well-being. Cognitively, AI challenges human exceptionalism, raising questions about non-human intelligence, consciousness, and the essence of being human. Governing AI also requires shifting from linear thinking to a more complex, systemic understanding of interconnected systems, where contradictions can coexist and reinforce each other.

### 4.2 Perspective of NGO Expert

#### 4.2.1 Structural Fault Lines in Global Governance

Global AI governance faces three structural contradictions: (1) Socio-technical Complexity: AI, as an invisible infrastructure, reshapes institutional and social relations, but national frameworks struggle to address the global impact, especially lacking strong normative guidance in areas such as human rights, inclusion, and labor protection within AI value chains [30]; (2) Paradox of Principles: The UN High-Level Advisory Body's (HLAB) four principles encounter practical barriers—the computing infrastructure gap hinders inclusivity (Principle 1), neo-colonial legal systems restrict the realization of the public interest (Principle 2), and data monopolies held by platform companies challenge sovereignty (Principle 3); (3) Institutional Path Dependence: The multi-stakeholder model has already shown problems of diluted consensus and power dominance in the Internet Governance Forum (IGF) (Principle 4).

#### 4.2.2 Reconstructing Multistakeholder Governance

Current governance models face a crisis of formalism: Southern nations cannot meaningfully participate due to the computing gap, and corporations dominate the agenda through "forum shopping." Reform requires three breakthroughs: (1) establishing digital public resource pools to ensure basic computing access for the Global South; (2) designing adversarial negotiation mechanisms to replace consensus principles with conflict management systems; (3) implementing corporate impact audits to mandate disclosure of technology transfer and commitments to local data management. The unresolved "enhanced cooperation" issue from the World Summit on the Information Society (WSIS) serves as a warning: participation without practical safeguards will ultimately become a theatrical exercise in governance.

#### 4.2.3 Pragmatic Transformations within Constraint

Given the slow pace of institutional change, a pragmatic approach focuses on three points: (1) adopting the BRICS model to establish a Southern AI innovation alliance, using public funds to support the development of foundational models; (2) forming issue-oriented lobbying groups within existing multilateral frameworks. (3) cultivating a public AI innovation track, which can also address the inherent challenge of interdisciplinary collaboration related to diversity, equity, and inclusion (DEI). These strategies are not substitutes for institutional change but rather aim to reshape the power landscape through cumulative practice.

### 4.3 HCI Perspective

#### 4.3.1 Theme 1: User-Driven Auditing and Decentralized AI Governance

This theme underscores the pivotal role of "everyday users" in algorithm auditing and value alignment [31, 32, 33, 34], illustrating how ordinary individuals can effectively identify and address biases in AI systems. Cases like the Twitter cropping algorithm, ImageNet Roulette, and Portrait AI demonstrate the impact of user-driven audits in exposing flaws and prompting corporate accountability. Such audits are characterized by two key features [35]: an ethics of care that prioritizes community building and diverse perspectives over abstract principles of justice, and a spontaneous, flexible nature that supports evolving, grassroots investigations. Users not only detect biases but also foster dialogue, build consensus, and mobilize collective action, highlighting how bottom-up initiatives can complement top-down regulatory efforts to promote decentralized, accountable AI governance. Achieving this requires strengthening public accountability mechanisms [36], fostering deliberative processes for conflict resolution [37, 38], and enhancing public AI literacy [39, 40].

#### 4.3.2 Theme 2: Co-Designing Responsible AI: A Participatory Approach to Early Intervention

This theme argues that many AI system failures stem from neglecting human and social factors during the design phase, rather than from technical flaws alone. To address this, it advocates for a participatory co-design approach that involves impact communities from the outset [41, 40]. By embedding community-centered processes that prioritize specific needs, values, and sustainability, this approach helps mitigate environmental and social risks, complementing user audits by addressing harms that may not be detectable post-deployment.

The success of both user-driven auditing and participatory co-design relies on several key factors. First, AI literacy is essential, as a basic understanding of AI concepts enables meaningful participation in both auditing and design processes [40]. Second, accessible platforms and tools, such as the AI life cycle comicboarding tool [41], are crucial for facilitating user input and collaboration. Finally, sufficient user engagement is necessary, as a critical mass of active participants is required to drive impactful change at organizational, legal, and policy levels.

#### 4.3.3 Theme 3: Essential Interdisciplinary Collaboration for Responsible AI

This theme highlights the critical need for interdisciplinary collaboration throughout all stages of AI development and governance. It argues that effective AI governance requires expertise beyond technical fields, including social scientists, ethicists, policymakers, and psychologists. This collaboration should begin at the task definition stage and continue through auditing and policy development to ensure socially responsible and ethically sound AI systems.

## 5 Interdisciplinary Convergence: Similarities, Differences, and Complementarities

This analysis compares and contrasts the perspectives of legal, NGO, and HCI experts on global AI governance, identifying key similarities, differences, and complementarities in the themes extracted from their speech. This approach reflects the core principles of interdisciplinary research, where collaboration among experts from different fields allows for the exploration of shared understandings and points of divergence, ultimately leading to the emergence of new knowledge [42],[25].

### 5.1 Dynamism (Adaptability and Responsiveness)

**Similarity**. All experts agree that AI governance must be dynamic to address rapidly changing technological and social environments.

**Differences**. Legal Expert: Focuses on institutional dynamism, advocating for flexible legal structures to address AI's global reach. NGO Expert: Highlights socio-political dynamism, emphasizing the need for constant adjustment in governance to address global inequalities and the unfinished agenda of the WSIS. HCI Expert: Promotes user-driven dynamism, emphasizing continuous feedback through audits and co-design, allowing AI systems to evolve based on user input.

**Complimentary**. These forms of dynamism are complementary. Institutional dynamism provides the overarching structure, socio-political dynamism ensures responsiveness to social needs and power dynamics, and user-driven dynamism grounds governance in practical experience and user feedback, creating a feedback loop for continuous improvement.

### 5.2 Experimentation (Iterative Development and Learning)

**Similarity**. All experts recognize that, given the novelty and complexity of AI technology, the governance process must incorporate elements of experimentation and learning.

**Differences**. Legal Expert: Advocates for regulatory experimentation, viewing AI as a catalyst for new governance models. NGO Expert: Focuses on policy experimentation, emphasizing incremental adjustments and South-South cooperation. HCI Expert: Highlights user-driven experimentation, where feedback from audits and co-design informs the iterative development of AI systems.

**Complementary**. These different forms of experimentation together constitute a more comprehensive and effective governance approach. Regulatory experimentation explores new governance models, policy experimentation provides practical experience and feedback, and user-driven experimentation ensures that AI systems remain aligned with user needs and ethical considerations.

### 5.3 Inclusivity (Participation and Representation)

**Similarity**. All experts agree on the importance of inclusivity in AI governance. They emphasize the need for broad participation and representation of diverse stakeholders in the decision-making process.

**Differences**. Legal Expert: Focuses on epistemic inclusivity, fostering interdisciplinary collaboration to build a shared understanding of AI's societal impact. NGO Expert: Stresses global inclusivity, calling for greater representation from the Global South and reforming multi-stakeholder models to address equity and power imbalances. HCI Expert: Promotes user inclusivity, integrating users directly into the AI development process to ensure systems reflect diverse needs and values.

**Complementary**. These different dimensions of inclusivity are complementary. Epistemic inclusivity provides the foundation for informed decision-making, global inclusivity ensures that all voices are heard, and user inclusivity grounds governance in the real-world experiences of those impacted by AI. This multi-pronged approach fosters a more comprehensive and equitable approach to AI governance.

### 5.4 Paradoxical Governance (Balancing Conflicting Objectives)

**Similarity**. All experts recognize that AI governance faces inherent contradictions and tensions, requiring complex governance strategies to address them. They all implicitly suggest the need for trade-offs and balances between different objectives.

**Differences**. Legal Expert: Identifies macro-level paradoxes, such as the need for global regulation within existing international institutions, which are limited by the "global race" mentality. NGO Expert: Focuses on meso-level tensions, where the pursuit of ideal goals, like a "hyperliberal" approach, conflicts with the constraints of neoliberal institutions. HCI Expert: Examines micro-level contradictions, such as the tension between user agency and the control exerted by tech companies, and the need for both early intervention and post-deployment auditing.

**Complementary**: These different levels of paradoxical governance are complementary. Legal experts focus on macro-level institutional and regulatory challenges. The proposed regulatory frameworks provide a macro-level institutional backdrop for the concerns of other experts. NGO experts focus on meso-level ideological conflicts within existing institutions, connecting macro structures to micro user agency and explaining the difficulty of achieving ideal goals. HCI experts examine micro-level user interactions and power dynamics with tech companies, grounding abstract concepts in lived experiences and showing how users navigate these issues. Integrating these macro (legal/institutional), meso (ideological/implementation), and micro (user/technology) analyses provides a more comprehensive understanding of AI governance's contradictions, building a richer framework for addressing its challenges.

# 6 Synergistic Multilevel Framework for Global AI Governance

The interdisciplinary convergence of legal, NGO, and HCI perspectives reveals key principles—dynamism, experimentation, inclusivity, and paradoxical governance—while highlighting the challenges of translating them into practice. To address this, we propose a synergistic multilevel framework that operationalizes these principles through concrete measures such as legal sandboxes, policy pilots, and participatory design workshops, derived from the complementary insights of our expert analysis.

## 6.1 Phase 1: Foundation Building

The first stage aims to establish an adaptive governance framework to address the core contradictions such as the accelerated pace of technological iteration and legal lag, global value divergence, and power imbalance. Legal experts design flexible regulatory sandboxes and cross-border contract templates to define legal experimental fields for rapidly evolving AI technologies; NGO experts simultaneously launch cross-cultural ethical dialogues to identify the special demands of the Global South in terms of data sovereignty and algorithmic justice; HCI experts conducted large-scale user perception research, mapping public AI awareness through focus groups and surveys. The synergy of the three enables the governance framework to maintain both institutional extensibility and embedded multi-cultural sensitivity and user value anchor points.

## 6.2 Phase 2: Experimental Verification

The second stage focuses on exposing practical contradictions such as policy efficiency and fairness, technological innovation and security, privacy protection and data utility. Legal experts contribute by overseeing regulatory sandboxes and piloting data governance frameworks (e.g., data trusts, data cooperatives) to explore different regulatory approaches and address legal uncertainties. NGO experts implement and monitor policy pilots, such as AI-driven public services in specific regions, to evaluate their social impact and address potential biases or inequalities, bridging the gap between policy design and real-world implementation. HCI experts design and conduct user-centered experiments, including A/B testing of different AI interfaces and user audits of algorithmic fairness, to gather empirical data on user experience and identify potential harms, providing valuable feedback for technology and policy refinement. This three-dimensional verification mechanism of 'legal sandbox - regional pilot - user experiment' transforms abstract governance principles into observable practical parameters.

## 6.3 Phase 3: Collaborative Optimization

The third stage committed to coordinating the conflicts of interest and resource gaps among multiple parties, and addressing the imbalance of technological power between the North and South and the cognitive barriers across disciplines HCI experts facilitate participatory design workshops and user forums to ensure diverse user needs and values are incorporated into AI development and governance. This ensures user voices are central to the process. NGO experts establish multi-stakeholder platforms for dialogue and negotiation, promoting global cooperation and addressing power imbalances between different actors, particularly between the Global North and South. This addresses the socio-political dynamics of AI governance. Legal experts develop international norms, standards, and legal frameworks based on the outcomes of pilots and stakeholder consultations, providing a more stable and predictable regulatory environment. This bottom-up value transmission mechanism ensures that the governance framework possesses both technical rationality and political legitimacy.

## 6.4 Phase 4: Global Adaptation

The fourth stage requires resolving the tension between standardization and localization of governance solutions. Legal experts develop mechanisms for international cooperation and harmonization of legal frameworks, addressing cross-border data flows, liability, and other transnational issues. This ensures global consistency while allowing for local adaptation. NGO experts facilitate knowledge sharing and capacity building across different regions, supporting the implementation of equitable and context-specific AI governance strategies. This bridges the gap between global norms and local practices. HCI experts develop tools and resources for local communities to adapt and implement user-centered AI governance practices, ensuring that global frameworks remain responsive to local needs. This empowers local communities to shape their own AI futures. The three levels of collaboration allow the global governance network to maintain the unity of core principles while allowing for innovation in local practices.

### 6.5 Phase 5: Continuous Evolution

The fifth stage mitigates the risk of institutional obsolescence due to technological disruptions by establishing a "monitor-analyze-iterate" learning loop. HCI experts establish ongoing user feedback loops and participatory evaluation processes to monitor the real-world impact of AI systems and governance frameworks. This provides continuous data for improvement. NGO experts conduct ongoing research and analysis of the social, economic, and ethical implications of AI, informing policy adjustments and promoting public discourse. This ensures governance remains responsive to evolving societal needs. Legal experts develop mechanisms for regular review and revision of legal frameworks, incorporating new knowledge and addressing emerging challenges. This ensures the legal framework remains adaptable to the evolving technological and social landscape. This dynamic mechanism, which links user behavior data, social cost analysis, and legal responses, enables the governance system to have the evolutionary resilience to cope with unknown challenges.

## 7 Discussion

Guided by the Integration and Implementation Sciences (IIS), this study conducted a participatory governance experiment at the United Nations Science Summit, fostering interdisciplinary dialogue among legal scholars, Global South NGO practitioners, and human-computer interaction (HCI) experts. Insights from domain experts were extracted through thematic analysis. An interdisciplinary comparative analysis identified four core principles of AI governance. These principles are operationalized through concrete measures, derived from the complementary insights of experts, and embedded into a five-phase, time-sequential model, forming a multilevel global AI governance framework.

The four core principles of global AI governance are: dynamism, experimentation, inclusivity and paradoxical governance. Dynamism, as an institutional gene, is explored through the differentiated interpretations of the three disciplines. Legal experts contribute flexible frameworks, NGOs offer value adjustments, and HCI provides user feedback, collectively verifying the applicability of Giddens' duality of structure theory in the digital age [43]. The legal sandbox, acting as a "structure," is "deconstructed" through South-South policy pilots and "restructured" via user feedback, forming a closed loop of institutional evolution. This process deepens March's exploration-exploitation balance theory [44], demonstrating that effective digital governance requires a triple dynamic synergy: institutional-level fault-tolerant design, organizational-level trial-and-error space, and individual-level error correction mechanisms.

Experimentation drives knowledge production through the complementary three-domain model of regulatory sandboxes, policy pilots, and user testing. This creates a novel knowledge production network where legal experiments generate normative knowledge (e.g., data trust compliance thresholds), NGO pilots produce contextual knowledge (e.g., cultural sensitivity parameters), and HCI tests yield experiential knowledge (e.g., privacy trade-off curves). This multi-modal knowledge symbiosis resonates with Nowotny's contextualized science paradigm [45], offering a knowledge transformation path for global governance that moves beyond simply replicating "best practices."

Inclusivity reconstructs governance legitimacy through a three-tiered mechanism encompassing cognitive integration, resource balance, and experience embedding. This approach overcomes the limitations of Habermas's deliberative democracy [46]. The participatory design in Phase 3 exemplifies how encoding user needs into micro-level technical parameters, translating them into meso-level resource agreements via North-South computing power negotiations, and ultimately shaping macro-level international responsibility standards, elevates governance legitimacy from "procedural compliance" to genuine "value symbiosis."

Paradoxical governance actively fosters system resilience by directly confronting the layered complexities of AI governance paradoxes. Legal experts at the macro level create adaptable institutional frameworks to manage the tension between rigid institutions and rapid technological change. At the meso level, NGOs establish adversarial negotiation platforms to navigate ideological differences. At the micro level, HCI experts use user audit tools to mitigate the conflict between corporate control and public oversight. This layered approach, incorporating institutional design, negotiation mechanisms, and technical tools, progressively resolves contradictions, validating Smith's paradox nesting theory [47].

The five-stage, multi-level governance framework unfolds the four core principles—dynamism, experimentation, inclusivity, and paradoxical governance—across time and space. Foundation Building (Phase 1) establishes institutional flexibility through legal sandboxes, creating a "safe experimental space" for interdisciplinary collaboration. Experimental Verification (Phase 2) tests the dynamism hypothesis via policy pilots and strengthens inclusivity through South-South cooperation. Collaborative Optimization (Phase 3) institutionalizes user demands as technical parameters through participatory design, reflecting the micro-level dynamism central to HCI. Global Adaptation (Phase 4) leverages compliance adaptors to achieve flexible expansion of the legal framework, demonstrating the transformative wisdom of paradoxical governance. Continuous Evolution (Phase 5) then ensures the ongoing balance of these four principles through a monitoring-analysis-iteration feedback loop. This "principles-guiding-stages, stages-solidifying-principles" cyclical structure surpasses the limitations of many multilateral initiatives, such as those from UNESCO, by providing concrete implementation pathways.

## 8 Contribution

This study makes three seminal contributions to AI governance research. Theoretically, it introduces the first integrative framework to addresses critical gaps in global AI governance, which currently lacks robust interdisciplinary collaboration, relies on symbolic multilateralism, and suffers from insufficient top-down and bottom-up cooperation.

Methodologically, this study pioneers the application of Integration and Implementation Sciences (IIS) to global AI governance, utilizing the UN summit as a participatory laboratory. By demonstrating how scientifically structured interdisciplinary dialogue can codify fragmented knowledge into an actionable five-stage framework, it offers a novel approach to translating complex challenges into concrete governance solutions.

Practically, this research offers practical strategies within the five-phases governance framework derived from the complementary of disciplinary insights. By providing these concrete implementation pathways, this research smoothly translates abstract governance principles into effective, practical management measures.

## 9 Limitation and Future Work

While strategic sampling provided depth in cross-disciplinary dynamics, the focused inclusion of experts from three particular areas and small sample size restrict generalizability—trade-offs necessary for prototyping novel governance interfaces. Future work should expand participant diversity and integrate additional disciplines like economics to test framework scalability. Though qualitative phase transitions proved effective for initial validation, developing quantitative maturity metrics could enhance operational precision. Crucially, this study demonstrates how structured dialogues transform disciplinary tensions into actionable frameworks—a replicable model for emerging fields like climate change governance. Subsequent iterations could scale the IIS "necessary diversity" principle while maintaining experimental rigor, positioning this work as the first evidence-based blueprint for translating multilateral debates into adaptive governance systems.

## 10 Conclusion

This study, through in-depth collaboration among legal, NGO, and HCI experts, constructs the first global AI governance framework that integrates interdisciplinary complementarity, multilateral consultation mechanisms, and vertical coordination paths. The multilevel governance framework design transforms dynamism, experimentation, inclusivity, and paradoxical governance into an operable evolutionary system, demonstrating that when institutional design accommodates disciplinary differences, negotiation platforms balance geopolitical power, and user participation drives continuous iteration, humanity is expected to achieve a paradigm shift in technology governance from crisis response to opportunity creation. This not only provides a new blueprint for AI regulation but also points out a common solution for the governance of complex global issues—cultivating synergy in differences and activating innovation in tension.

## 11 Acknowledgment

We wish to acknowledge and thank Ms. Nandini Chami, the deputy director of IT for Change, for her generous contribution to the UN Science Summit as a guest speaker. Her presentation, offering a vital Global South perspective on AI governance, provided invaluable insights and enriched our understanding of this complex issue.

Global AI Governance: Where the Challenge is the Solution- An Interdisciplinary, Multilateral, and Vertically Coordinated Approach